\documentclass[aip, pop, amsfonts, amssymb, amsmath ,reprint,subscriptadress]{revtex4-1}
\usepackage{amssymb}		 
\usepackage{amsmath, amsthm, amssymb,amsfonts}
\usepackage[english]{babel}
\usepackage{color, graphicx}
\usepackage{epstopdf} 
\usepackage{caption}
\usepackage{comment}
\usepackage{hyperref}
\begin{document}

\title{Collision-dependent power law scalings in 2D gyrokinetic turbulence}
%\author{S. S. Cerri$^1$}\email{silvio.sergio.cerri@ipp.mpg.de}
%\author{A. Ba\~n\'on Navarro$^1$}
%\author{F. Jenko$^{1,2}$}
%\author{D. Told$^1$}
%\affiliation{$^1$Max-Planck-Institut f\"ur Plasmaphysik, Boltzmannstr. 2, D-85748 Garching, Germany}
%\affiliation{$^2$Max-Planck/Princeton Center for Plasma Physics}
\author{S. S. Cerri}\email{silvio.sergio.cerri@ipp.mpg.de}\affiliation{Max-Planck-Institut f\"ur Plasmaphysik, Boltzmannstr. 2, D-85748 Garching, Germany}
\author{A. Ba\~n\'on Navarro}\affiliation{Max-Planck-Institut f\"ur Plasmaphysik, Boltzmannstr. 2, D-85748 Garching, Germany}
\author{F. Jenko}\affiliation{Max-Planck-Institut f\"ur Plasmaphysik, Boltzmannstr. 2, D-85748 Garching, Germany}\affiliation{Max-Planck/Princeton Center for Plasma Physics}
\author{D. Told}\affiliation{Max-Planck-Institut f\"ur Plasmaphysik, Boltzmannstr. 2, D-85748 Garching, Germany}

%**************%
%   ABSTRACT   %
%**************%
\begin{abstract}

Nonlinear gyrokinetics provides a suitable framework to describe short-wavelength turbulence in magnetized laboratory and astrophysical plasmas. In the electrostatic limit, this system is known to exhibit a free energy cascade towards small scales in (perpendicular) real and/or velocity space. The dissipation of free energy is always due to collisions (no matter how weak the collisionality), but may be spread out across a wide range of scales. Here, we focus on freely-decaying 2D electrostatic turbulence on sub-ion-gyroradius scales. An existing scaling theory for the turbulent cascade in the weakly collisional limit is generalized to the moderately collisional regime. In this context, non-universal power law scalings due to multiscale dissipation are predicted, and this prediction is confirmed by means of direct numerical simulations.
\end{abstract}

%\pacs{52.25.Xz, 52.25.Dg, 95.30.Qd}

\maketitle

%******************%
%   INTRODUCTION   %
%******************%
\section{Introduction}\label{sec:intro}

Plasma turbulence in weakly collisional magnetized plasmas plays an important role in various systems, such as fusion devices and many space and astrophysical situations, where it leads, e.g., to anomalous transport effects and particle heating. Such plasmas are usually almost collisionless and thus the turbulence problem requires a kinetic approach, especially at small scales where dissipation takes place. In the standard picture, turbulence can be interpreted as a conservative transfer of energy in wavenumber space, from injection to dissipation scales~\citep{Frisch1995,FalkovichPT2006}. While the fundamental processes in the hydrodynamic case, represented by the Navier-Stokes equation, are fairly well understood at this point, the turbulence theory of magnetized plasmas is still far from complete. Over the last several years, it has become clear that plasma microturbulence - as described by nonlinear gyrokinetic (GK) theory~\citep{FriemannPOF1982,BrizardRMP2007,KrommesARFM2012} - cannot be viewed as a straightforward extension of fluid turbulence.

In the 3D Navier-Stokes system, the {\em kinetic energy}, assumed to be injected into the system at large scales through mechanical forces, is conserved by the advective nonlinearity which is responsible for transferring the energy from the injection scales to the smallest ones (turbulent {\em cascade}) at which the energy is then dissipated by viscous effects~\citep{Frisch1995,FalkovichPT2006}. In the GK formalism, on the other hand, the ideal quadratic invariant is given by the {\em free energy} which is subject to a phase-space cascade towards small scales in (perpendicular) real and/or velocity space~\citep{SchekochihinAPJS2009,BanonNavarroPRL2011}. A Kolmogorov-like phenomenological scaling theory of GK turbulence has been developed for the weakly collisional limit~\citep{SchekochihinPPCF2008,PlunkJFM2010}. The respective predictions have also been confirmed via direct numerical simulations~\citep{TatsunoPRL2009,TatsunoJPFR2010}. The main goal of the present work is to extend this work to the moderately collisional regime, focusing, for simplicity, on freely decaying 2D electrostatic turbulence on sub-ion-gyroradius scales. As it will turn out, our generalization predicts non-universal power law scalings due to multiscale dissipation, and this prediction is confirmed by means of direct numerical simulations.

The remainder of this paper is organized as follows. An introduction to the GK system of equations in the 2D electrostatic limit and to its global and local energy balance equations is given in Sec.~\ref{sec:system}. Then, in Sec.~\ref{sec:scalings} we provide a brief review of the theory of nonlinear phase mixing at sub-ion-gyroradius scales as proposed in Refs.~\citep{SchekochihinPPCF2008,PlunkJFM2010} (Sec.~\ref{subsec:theory_0}), and we propose a novel natural extension of such theory to the multiscale dissipation case (Sec.~\ref{subsec:theory_new}). In Sec.~\ref{sec:sims}, we present the results from direct numerical simulations with the GK plasma turbulence code GENE~\citep{JenkoPOP2000,DannertPOP2005,GoerlerJCP2011}. We demonstrate the validity of the assumptions made in the theory we developed in Sec.~\ref{subsec:theory_new} for increasing collisionality by comparisons with the numerical results. Moreover, in the low collisionality limit, we recover the hypothesis and the results predicted by the standard theory of Refs.\citep{SchekochihinPPCF2008,PlunkJFM2010}, showing that the transition from one case to the other is continuous, as one might have expected.

%*************************%
%   SYSTEM OF EQUATIONS   %
%*************************%
\section{The system under study}\label{sec:system}

In the present study, the full GK system of equations is reduced to the simple scenario of a single ion species, electrostatic perturbations, slab geometry with $k_\|=0$ (thus avoiding parallel effects such as the Landau damping and the linear phase mixing~\citep{LandauJP_1946,HammettPFB-4_1992} to be effective), and no-response electrons (which is consistent with $k_\|=0$). No gradients in the background quantities (i.e., the density $n_0$, the temperature $T_0$, and the magnetic field $B_0$) are considered. This allows us to focus on both nonlinear and collisional effects, which are the features we are primarily interested in. We are considering the total ion distribution function $F$ to be split into a Maxwellian part, $F_0=(n_0/\pi^{3/2}v_T^3)\exp(-v^2/v_T^2)$, $v_T=\sqrt{2T_0/m}$ being the thermal velocity, and a perturbed part, $F_1$, i.e., $F=F_0+F_1$. Then, the gyrokinetic Vlasov equation for the perturbed distribution function $F_1$ and the gyrokinetic Poisson equation for the self-consistent electrostatic potential $\phi_1$ read 
%%%%%%%%%%%%%%%%%%%%%%%%%%%%%%%%%%%%%%%%%%%%%%%%%%%%%
\begin{equation}\label{eq:GK-Vlasov}
\frac{\partial F_1}{\partial t} + {\bf v}_{\bar{\phi}}\cdot\nabla F_1\ =\ \langle C^L[F_1]\rangle
\end{equation}
and
%%%%%%%%%%%%%%%%%%%%%%%%%%%%%%%%%%%%%%%%%%%%%%%%%%%%%
%%%%%%%%%%%%%%%%%%%%%%%%%%%%%%%%%%%%%%%%%%%%%%%%%%%%%
\begin{equation}\label{eq:GK-Poisson}
 \frac{en_0}{T_0}\big(1-\Gamma_0\big)\phi_1\ =\ \frac{2\pi B_0}{m}\int J_0(k_\perp\rho)F_1{\rm d}v_\|{\rm d}\mu\,,
\end{equation}
%%%%%%%%%%%%%%%%%%%%%%%%%%%%%%%%%%%%%%%%%%%%%%%%%%%%%
where ${\bf v}_{\bar{\phi}}=(c/B_0)({\bf e}_z\times\nabla\bar{\phi}_1)$ is the E$\times$B drift due to the gyroaveraged self-consistent electrostatic potential $\bar{\phi}_1$, the background magnetic field being aligned with the $z$-axis, $\langle C^L[F_1]\rangle$ is a linearized collision operator, $\Gamma_0\ \equiv\ \frac{2\pi B_0}{mn_0}\int J_0^2(k_\perp\rho)F_0dv_\|d\mu$, $J_0$ is the Bessel function, $\rho=v_\perp/\Omega_c$ is the Larmor radius, and $\mu=mv_\perp^2/2B_0$ is the magnetic moment. Note that, in the periodic case we are using, the gyroaverage can be written as a multiplication by the Bessel function $J_0$, e.g., $\bar{\phi}_1=J_0\phi_1$. The term ${\bf v}_{\bar{\phi}}\cdot\nabla F_1$ is the nonlinear term and, because of the gyroaverage of the electrostatic potential, it is responsible for the {\em nonlinear phase mixing} process\citep{SchekochihinPPCF2008,PlunkJFM2010}.

In the absence of collisions, due to the conservative property of the nonlinear term, Eqs.(\ref{eq:GK-Vlasov})-(\ref{eq:GK-Poisson}) admit two positive definite conserved integrals~\citep{SchekochihinPPCF2008,PlunkJFM2010,BanonNavarroPOP2011}. One of them is quadratic in $F_1$, which is then proportional to (minus) the perturbed part of the entropy (of the gyrocenters), while the other one is proportional to the product of $\bar{\phi}_1$ and $F_1$, which is usually referred to as the electrostatic energy (or polarization term):
%%%%%%%%%%%%%%%%%%%%%%%%%%%%%%%%%%%%%%%%%%%%%%%%%%%%%
\begin{equation}\label{eq:entropy}
\mathcal{E}_f = \int\frac{T_0F_1^2}{2F_0}{\rm d}\Lambda
\end{equation}
%%%%%%%%%%%%%%%%%%%%%%%%%%%%%%%%%%%%%%%%%%%%%%%%%%%%%
and
%%%%%%%%%%%%%%%%%%%%%%%%%%%%%%%%%%%%%%%%%%%%%%%%%%%%%
\begin{equation}\label{eq:es_energy}
\mathcal{E}_\phi = \int\frac{e\bar{\phi}_1F_1}{2}{\rm d}\Lambda\,,
\end{equation}
%%%%%%%%%%%%%%%%%%%%%%%%%%%%%%%%%%%%%%%%%%%%%%%%%%%%%
where ${\rm d}\Lambda\equiv{\rm d}x\,{\rm d}y\,{\rm d}\Theta = \pi n_0B_0\,{\rm d}x\,{\rm d}y\,{\rm d}v_\|\,{\rm d}\mu$ is the phase-space element. Note that, using the Poisson equation and the local approximation, the electrostatic energy can also be written as
%%%%%%%%%%%%%%%%%%%%%%%%%%%%%%%%%%%%%%%%%%%%%%%%%%%%%
\begin{equation}
{\cal E}_\phi=\int{\rm d}x\,{\rm d}y\,(1-\Gamma_0)\phi_1^2\ .
\end{equation}
%%%%%%%%%%%%%%%%%%%%%%%%%%%%%%%%%%%%%%%%%%%%%%%%%%%%%
The entropy and the electrostatic energy, together, define the free energy:
%%%%%%%%%%%%%%%%%%%%%%%%%%%%%%%%%%%%%%%%%%%%%%%%%%%%%
\begin{equation}
{\cal E} = {\cal E}_f + {\cal E}_\phi\,.
\end{equation}
%%%%%%%%%%%%%%%%%%%%%%%%%%%%%%%%%%%%%%%%%%%%%%%%%%%%%
However, when collisions are taken into account, the global energy balance equations read
%%%%%%%%%%%%%%%%%%%%%%%%%%%%%%%%%%%%%%%%%%%%%%%%%%%%%
%\begin{subequations} 
%\label{eq:glob_energy}
%\begin{equation}
%\frac{\partial{\cal E}_f}{\partial t} = - {\cal C}_f - {\cal H}_f
%\,,
%\end{equation}
%\begin{equation}
%\frac{\partial{\cal E}_\phi}{\partial t} = - {\cal C}_\phi - {\cal H}_\phi
%\,,
%\end{equation}
%\end{subequations}
%%%%%%%%%%%%%%%%%%%%%%%%%%%%%%%%%%%%%%%%%%%%%%%%%%%%%
%%%%%%%%%%%%%%%%%%%%%%%%%%%%%%%%%%%%%%%%%%%%%%%%%%%%%
\begin{equation}\label{eq:glob_energy}
 \frac{\partial{\cal E}_{\{f,\phi\}}}{\partial t} = - {\cal C}_{\{f,\phi\}} - {\cal H}_{\{f,\phi\}}\,,
\end{equation}
%%%%%%%%%%%%%%%%%%%%%%%%%%%%%%%%%%%%%%%%%%%%%%%%%%%%%
where the collisional dissipation terms are
%%%%%%%%%%%%%%%%%%%%%%%%%%%%%%%%%%%%%%%%%%%%%%%%%%%%%
\[ {\cal C}_f = -\int\frac{T_0}{F_0}F_1\langle C^L[F_1]\rangle\,{\rm d}\Lambda\,, \]
\[ {\cal C}_\phi = -\int e\bar{\phi}_1\langle C^L[F_1]\rangle\,{\rm d}\Lambda\,, \]
%%%%%%%%%%%%%%%%%%%%%%%%%%%%%%%%%%%%%%%%%%%%%%%%%%%%%
while ${\cal H}_f$ and ${\cal H}_\phi$ are extra dissipation terms due to $k_\perp$-hyperdiffusion, which will be defined later and will be negligible for the $k_\perp$-range of interest (see Sec.~\ref{sec:sims}).

Going to the local energy balance equations, in the Fourier representation they are
%%%%%%%%%%%%%%%%%%%%%%%%%%%%%%%%%%%%%%%%%%%%%%%%%%%%%
\begin{subequations} 
\label{eq:loc_energy}
\begin{equation}
\frac{\partial E_f(k)}{\partial t} = T_f(k) - C_f(k) - D_{\perp,f}(k)
\end{equation}
\begin{equation}
\frac{\partial E_\phi(k)}{\partial t} =  T_\phi(k) - C_\phi(k) - D_{\perp,\phi}(k)\,,
\end{equation}
\end{subequations}
%%%%%%%%%%%%%%%%%%%%%%%%%%%%%%%%%%%%%%%%%%%%%%%%%%%%%
where the spectral density of entropy $E_f(k)$ is defined by
%%%%%%%%%%%%%%%%%%%%%%%%%%%%%%%%%%%%%%%%%%%%%%%%%%%%%
\[ {\cal E}_f = \int{\rm d}x{\rm d}y\int\frac{T_0F_1^2}{2F_0}{\rm d}\Theta \]
\[ = \sum_k \int\frac{T_0|f_k|^2}{2F_0}{\rm d}\Theta = \sum_k E_f(k)\,, \]
%%%%%%%%%%%%%%%%%%%%%%%%%%%%%%%%%%%%%%%%%%%%%%%%%%%%%
where the sum is over all the $k_x$ and $k_y$. Equivalently, we define the spectral density of the entropy collisional and perpendicular dissipations, i.e. ${\cal C}_f\equiv\sum_k C_f(k)$ and ${\cal H}_f\equiv\sum_k D_{\perp,f}(k)$, respectively. The nonlinear transfer of entropy $T_f(k)$ is given by~\citep{BanonNavarroPOP2011} 
%%%%%%%%%%%%%%%%%%%%%%%%%%%%%%%%%%%%%%%%%%%%%%%%%%%%%
\[ T_f(k) = \sum_{k'}{\cal T}_f(k,k') \]
\[ =\sum_{k'}\int{\rm d}\Theta\frac{T_0}{F_0}f_k^*\big[(k_x-k_x')\bar{\phi}_{1(k-k')}k_y'f_{k'}\]
\[ -(k_y-k_y')\bar{\phi}_{1(k-k')}k_x'f_{k'}\big]\,, \]
%%%%%%%%%%%%%%%%%%%%%%%%%%%%%%%%%%%%%%%%%%%%%%%%%%%%%
where $f_k^*$ is the complex conjugate of $f_k$ and the sum is over all the $k_x'$ and $k_y'$. In a similar way, we define a nonlinear transfer term and a spectral density of the electrostatic energy, $T_\phi(k)$ and $E_\phi(k)$, respectively. 
Note that the nonlinear term is the only term responsible for a transfer of entropy or of electrostatic energy between different Fourier modes, e.g. between $f_{k'}$ and $f_k$ due to $\bar{\phi}_{1(k-k')}$ (which is linear in $f_{k-k'}$ because of the Poisson equation in Fourier space). This determines a so-called triadic interaction between the modes $f_k$, $f_{k-k'}$ and $f_{k'}$, which is interpreted as an exchange of energy between modes $k$ and $k'$ due to the property ${\cal T}_f(k,k')=-{\cal T}_f(k',k)$ (from which follows the conservative behavior of this term, i.e. $\sum_kT(k)=\sum_{k,k'}{\cal T}(k,k')=-\sum_{k,k'}{\cal T}(k',k)=0$). 

Usually, the nonlinear transfer $T(k)$ is interpreted as (minus) the divergence of a flux in wavenumber space, i.e.
%%%%%%%%%%%%%%%%%%%%%%%%%%%%%%%%%%%%%%%%%%%%%%%%%%%%%
\[ T(k) = - \frac{\partial\Pi(k)}{\partial k}\,, \]
%%%%%%%%%%%%%%%%%%%%%%%%%%%%%%%%%%%%%%%%%%%%%%%%%%%%%
which means that the local energy balance equation has the form
%%%%%%%%%%%%%%%%%%%%%%%%%%%%%%%%%%%%%%%%%%%%%%%%%%%%%
\begin{equation}\label{eq:local_energy_2}
 \frac{\partial E(k)}{\partial t} + \frac{\partial\Pi(k)}{\partial k} = - D_{\rm tot}(k)\,,
\end{equation}
%%%%%%%%%%%%%%%%%%%%%%%%%%%%%%%%%%%%%%%%%%%%%%%%%%%%%
where now $D_{\rm tot}$ represents all the possible dissipation sources. Then, one usually assumes a quasi-stationary state (so $\partial_t E(k)\approx0$ after a time-average) and considers the inertial range ($D_{\rm tot}(k)\approx0$), so the above balance equation reduces to
%%%%%%%%%%%%%%%%%%%%%%%%%%%%%%%%%%%%%%%%%%%%%%%%%%%%%
\[ \frac{\partial\Pi(k)}{\partial k} = 0\quad\Rightarrow\quad \Pi(k)=\Pi_0={\rm const.}\,, \]
%%%%%%%%%%%%%%%%%%%%%%%%%%%%%%%%%%%%%%%%%%%%%%%%%%%%%
and it can be solved with the help of a closure relation between the flux and the spectrum, if such a relation exists, e.g. 
%%%%%%%%%%%%%%%%%%%%%%%%%%%%%%%%%%%%%%%%%%%%%%%%%%%%%
\[ \Pi(k)\sim k^\alpha E(k) \quad\Rightarrow\quad E(k)\propto k^{-\alpha}\,. \]
%%%%%%%%%%%%%%%%%%%%%%%%%%%%%%%%%%%%%%%%%%%%%%%%%%%%%
However, these arguments apply if the turbulence can be assumed to be in a quasi-stationary state (which is usually the case for driven turbulence) and if there exists a range over which the dissipation and the drive is negligible (i.e., what is commonly called inertial range). Unfortunately, this seems not to be the case for freely-decaying sub-Larmor scale turbulence, which is what we are going to study in the present work. Thus a more general approach based on heuristic and physical arguments is required.

%**************%
%   SCALINGS   %
%**************%
\section{Scalings and collisions}\label{sec:scalings}

For the almost collisionless (or weakly collisional) case, a scaling theory of the entropy cascade in sub-Larmor scale range has been proposed~\citep{SchekochihinPPCF2008,PlunkJFM2010} and it has been tested by means of direct numerical simulations~\citep{TatsunoPRL2009,TatsunoJPFR2010}.
However, such Kolmogorov-style arguments are based on several assumptions, one of which is the existence of an inertial range in which no dissipation occurs. 
Nevertheless, gyrokinetics can exhibit multiscale dissipation throughout a wide wave number range, as it was shown in recent papers~\citep{TeacaPRL2012,HatchPRL2013,TeacaPOP2014}.
With the present work, we want to add a new, simple system to the previously mentioned cases, showing that even in our case multiscale dissipation is present and, moreover, it can also affect the spectra exponents, leading to non-universal power laws. Nonetheless, this work confirms again the weakly collisional theory in the proper limit, extending it in the moderate collisionality limit (i.e., where the spectra still have a significant power law component and they are not just an exponential fall off, which would be the case in the very high collisionality limit).

\subsection{A review of the weakly collisional theory}\label{subsec:theory_0}

Before going to the intermediate collisionality case, let us quickly review the weakly collisional theory suggested in Refs.~\citep{SchekochihinPPCF2008,PlunkJFM2010}. In our simple system, the only term responsible for the energy transfer among modes is the nonlinear term in Eq.~(\ref{eq:GK-Vlasov}), ${\bf v}_{\bar{\phi}}\cdot\nabla F_1$, from which we can readily estimate the nonlinear decorrelation rate $\omega_{NL}$, i.e.,
%%%%%%%%%%%%%%%%%%%%%%%%%%%%%%%%%%%%%%%%%%%%%%%%%%%%%
\begin{equation}\label{eq:NLrate_0}
\omega_{NL} \sim k_\perp J_0(k_\perp\rho)\phi_kk_\perp \sim k_\perp^{3/2}\phi_k\,,
\end{equation}
%%%%%%%%%%%%%%%%%%%%%%%%%%%%%%%%%%%%%%%%%%%%%%%%%%%%% 
where we have used the large argument approximation of the Bessel function, $J_0(\zeta)\approx\zeta^{-1/2}\cos(\zeta-\pi/4)$ for $\zeta\gg1$, since we are interested in the sub-ion-gyroradius scale range cascade, $k_\perp\rho\gg1$.
Then, since our aim is to first estimate the free energy flux, $\omega_{NL}f_k^2\sim k_\perp^{3/2}\phi_kf_k^2$, the next step is to relate the electrostatic potential components $\phi_k$ to those of the perturbed part of the distribution function $f_k$. This can be done through the GK Poisson equation, Eq.~(\ref{eq:GK-Poisson}); in the Fourier representation, it reads
%%%%%%%%%%%%%%%%%%%%%%%%%%%%%%%%%%%%%%%%%%%%%%%%%%%%%
\begin{equation}\label{eq:GK-Poisson_k}
 \phi_k = \widetilde{\beta}(k_\perp)\int v_\perp J_0\Big(\frac{k_\perp v_\perp}{\Omega}\Big)\hat{F}_1{\rm d}v_\|{\rm d}v_\perp\,,
\end{equation}
%%%%%%%%%%%%%%%%%%%%%%%%%%%%%%%%%%%%%%%%%%%%%%%%%%%%%
where $\widetilde{\beta}(k)=2\pi/(1-\hat{\Gamma}_0(k))=2\pi/(1-I_0(k^2)e^{-k^2})$, $I_0$ being the modified Bessel function. In the $k_\perp\rho\gg1$ limit, $\widetilde{\beta}(k)\approx{\rm const.}$ and the large argument expansion of the Bessel function $J_0$ gives
%%%%%%%%%%%%%%%%%%%%%%%%%%%%%%%%%%%%%%%%%%%%%%%%%%%%%
\begin{equation}\label{eq:GK-Poisson_k2}
 \phi_k \sim k_\perp^{-1/2}\int v_\perp^{-1/2} \cos\Big(\frac{k_\perp v_\perp}{\Omega}-\frac{\pi}{4}\Big)\hat{F}_1{\rm d}v_\|{\rm d}v_\perp\,.
\end{equation}
%%%%%%%%%%%%%%%%%%%%%%%%%%%%%%%%%%%%%%%%%%%%%%%%%%%%%
Now the nonlinear phase mixing argument comes into play and helps us to estimate the above integral by assuming the correspondence of length scales in real space, $l$, and in perpendicular velocity space\citep{lv_note}, $l_v$, due to the nonlinear phase mixing process, i.e., $l_v\sim l$. In fact, let's assume that $l$ is the correlation length of the E$\times$B flow (i.e., of the electrostatic potential $\phi_1$). Since in gyrokinetics the particle drift is determined by the fluctuating fields averaged over their gyro-orbits, particles sharing the same gyrocenter position, but having different perpendicular velocities $v_\perp$, are gyro-averaging over different Larmor orbits and thus they are experiencing different drifts in real space. If the difference in the Larmor radii, and thus in $v_\perp\propto\rho$, of such orbits is of the order of the correlation length $l$ of the averaged potential, then the two particles are decorrelated and they will perform independent random walks. This, in turns, mean that the distribution function $F_1$ develops random structures in $v_\perp$-space on the scales $l_v\sim l$ (see Refs.~\citep{SchekochihinPPCF2008,PlunkJFM2010} for further details). Then, because of the random nature this process, the velocity space integral (\ref{eq:GK-Poisson_k2}) can be argued to accumulate like a random walk in which the step size scales as $l_v$ and the number of steps scale as $l_v^{-1}$, so the typical displacement scales as $l_v^{1/2}$. In terms of $v_\perp$-scale conjugate variable $p$ ($p$ is for the Hankel transform in $v_\perp$-space what $k_\perp$ is for the Fourier transform in real space), this means
%%%%%%%%%%%%%%%%%%%%%%%%%%%%%%%%%%%%%%%%%%%%%%%%%%%%%
\begin{equation}\label{eq:phi-f_scal_0}
 \phi_k \sim k_\perp^{-1/2}p^{-1/2}f_k \sim k_\perp^{-1}f_k\,,
\end{equation}
%%%%%%%%%%%%%%%%%%%%%%%%%%%%%%%%%%%%%%%%%%%%%%%%%%%%%
where the last step comes from the nonlinear phase mixing argument $l\sim l_v$ and $f_k$ has to be interpreted~\citep{PlunkJFM2010} as the root-mean-square value of the fluctuations in $\hat{F}_1$, i.e., $f_k\sim\sqrt{\int\hat{F}_1^2v_\perp{\rm d}v_\perp{\rm d}v_\|}$.
Now, from Eqs.~(\ref{eq:NLrate_0}) and (\ref{eq:phi-f_scal_0}), assuming locality of interactions and constancy of the flux of free energy $\Pi_f$, we obtain
%%%%%%%%%%%%%%%%%%%%%%%%%%%%%%%%%%%%%%%%%%%%%%%%%%%%%
\begin{equation}\label{eq:const_flux}
  \omega_{NL}f_k^2\sim k_\perp^{3/2}\phi_kf_k^2 \sim k_\perp^{-1/2}f_k^3 \sim \varepsilon_0 = {\rm const.}\,,
\end{equation}
%%%%%%%%%%%%%%%%%%%%%%%%%%%%%%%%%%%%%%%%%%%%%%%%%%%%%
from which we readily get $f_k\sim\varepsilon_0^{1/3}k_\perp^{-1/6}$ and $\phi_k\sim\varepsilon_0^{1/3}k_\perp^{-7/6}$ and finally the spectra:
%%%%%%%%%%%%%%%%%%%%%%%%%%%%%%%%%%%%%%%%%%%%%%%%%%%%% 
\begin{subequations} 
\label{eq:spectra_0}
\begin{eqnarray}
 E_f(k_\perp) & \sim & \varepsilon_0^{2/3}k_\perp^{-4/3}\\
 E_\phi(k_\perp) & \sim & \varepsilon_0^{2/3}k_\perp^{-10/3}\,.
\end{eqnarray}
\end{subequations}
%%%%%%%%%%%%%%%%%%%%%%%%%%%%%%%%%%%%%%%%%%%%%%%%%%%%% 
Note that in the above assumptions, two features can be taken as the main ones, i.e. the nonlinear phase mixing process (i.e., $l\sim l_v$) and the existence of a Kolmogorov-like inertial range over which absolutely no dissipation occurs (i.e., the constancy of the free energy flux $\Pi_f\sim\varepsilon_0={\rm const}$). In the following, we are going to relax the latter one and to slightly modify the argument from which we are estimating the velocity integral in Eq.(\ref{eq:GK-Poisson_k2}), while still assuming the nonlinear phase mixing argument $l\sim l_v$ to hold.

\subsection{Taking into account multiscale dissipation}\label{subsec:theory_new}

We now extend the above theory in order to take into account collisional effects.
First of all, let's assume that the nonlinear phase mixing argument, $l_v\sim l$, is still valid and that the decorrelation rate is still given by the nonlinear term, $\omega_{NL}\sim k_\perp^{3/2}\phi_k$, which is responsible for the conservative energy transfer in wavenumber space.
Then, with respect to the weakly collisional case, we argue that the scaling between $\phi_k$ and $f_k$ becomes steeper due to the collisional smearing of the very small velocity-space scales, i.e.,
%%%%%%%%%%%%%%%%%%%%%%%%%%%%%%%%%%%%%%%%%%%%%%%%%%%%%
\begin{equation}\label{eq:phi-f_scal}
\phi_k \sim k_\perp^{-1-\theta_\nu}f_k\,,
\end{equation}
%%%%%%%%%%%%%%%%%%%%%%%%%%%%%%%%%%%%%%%%%%%%%%%%%%%%% 
which is estimated from Eq.~(\ref{eq:GK-Poisson_k2}) with the same argument by which the velocity integral accumulates like a random walk, but now the step is a bit larger than in the weakly collisional case, an effect which is represented here via the $\theta_\nu$ correction.
Again, we now have to consider the flux of free energy, i.e. $\omega_{NL}f_k^2\sim k_\perp^{3/2}\phi_kf_k^2$, together with the scaling in Eq.~(\ref{eq:phi-f_scal}). In this regard, we do not assume a constancy of such flux as in the weakly collisional case (i.e., the existence of an inertial range over which no dissipation occurs), but we allow for a $k_\perp$-dependent $\Pi_f(k_\perp)$ due to multiscale dissipation:
%%%%%%%%%%%%%%%%%%%%%%%%%%%%%%%%%%%%%%%%%%%%%%%%%%%%%
\begin{equation}\label{eq:flux}
  \omega_{NL}f_k^2\sim k_\perp^{3/2}\phi_kf_k^2 \sim k_\perp^{-1/2}f_k^3 \sim \Pi_f(k_\perp)\,.
\end{equation}
%%%%%%%%%%%%%%%%%%%%%%%%%%%%%%%%%%%%%%%%%%%%%%%%%%%%%
In general, to take into account the dissipation cut-off, the flux can be expressed as a combination of a power law and an exponential, e.g.,
%%%%%%%%%%%%%%%%%%%%%%%%%%%%%%%%%%%%%%%%%%%%%%%%%%%%%
\[
 \Pi_f(k_\perp) \sim \varepsilon_0k_\perp^{-\delta_\nu}e^{-\beta k_\perp^\gamma}\,.
\]
%%%%%%%%%%%%%%%%%%%%%%%%%%%%%%%%%%%%%%%%%%%%%%%%%%%%%
Note that, in principle also the exponential parameters depend on the collisionality, i.e., $\beta=\beta_\nu$ and $\gamma=\gamma_\nu$. However, we assume that there exists a $k_\perp$-range over which the flux it is nearly a power law, i.e.,
%%%%%%%%%%%%%%%%%%%%%%%%%%%%%%%%%%%%%%%%%%%%%%%%%%%%%
\begin{equation}\label{eq:PIf_hp}
\Pi_f(k_\perp) \approx \varepsilon_0k_\perp^{-\delta_\nu}\,.
\end{equation}
%%%%%%%%%%%%%%%%%%%%%%%%%%%%%%%%%%%%%%%%%%%%%%%%%%%%% 
This leads to the following spectra:
%%%%%%%%%%%%%%%%%%%%%%%%%%%%%%%%%%%%%%%%%%%%%%%%%%%%% 
\begin{subequations} 
\label{eq:spectra}
\begin{eqnarray}
 E_f(k_\perp) & \sim & \varepsilon_0^{2/3}k_\perp^{-4/3-2(\delta_\nu-\theta_\nu)/3}\\
 E_\phi(k_\perp) & \sim & \varepsilon_0^{2/3}k_\perp^{-10/3-2(\delta_\nu+2\theta_\nu)/3}\,,
\end{eqnarray}
\end{subequations}
%%%%%%%%%%%%%%%%%%%%%%%%%%%%%%%%%%%%%%%%%%%%%%%%%%%%% 
which are in general steeper than the weakly collisional spectra in Eq.~(\ref{eq:spectra_0}), which are however consistently recovered by the limit $\theta_\nu$, $\delta_\nu\to0$. We finally note that, if we retain the exponential cut-off in the free energy flux $\Pi_f(k_\perp)$, we obtain the same result given in Eq.~(\ref{eq:spectra}), just multiplied by the exponential cut-off $\exp(-\frac{2}{3}\beta_\nu k_\perp^{\gamma_\nu})$.

Hereafter, we will refer to the spectra exponents as $\alpha_f$ and $\alpha_\phi$, such that $E_f(k_\perp)\propto k_\perp^{-\alpha_f}$ and $E_\phi(k_\perp)\propto k_\perp^{-\alpha_\phi}$.

%*****************%
%   SIMULATIONS   %
%*****************%
\section{Direct numerical simulations}\label{sec:sims}

\begin{figure}[!h]
  \begin{minipage}[!h]{0.2385\textwidth}
  \flushleft\includegraphics[width=1.0\textwidth]{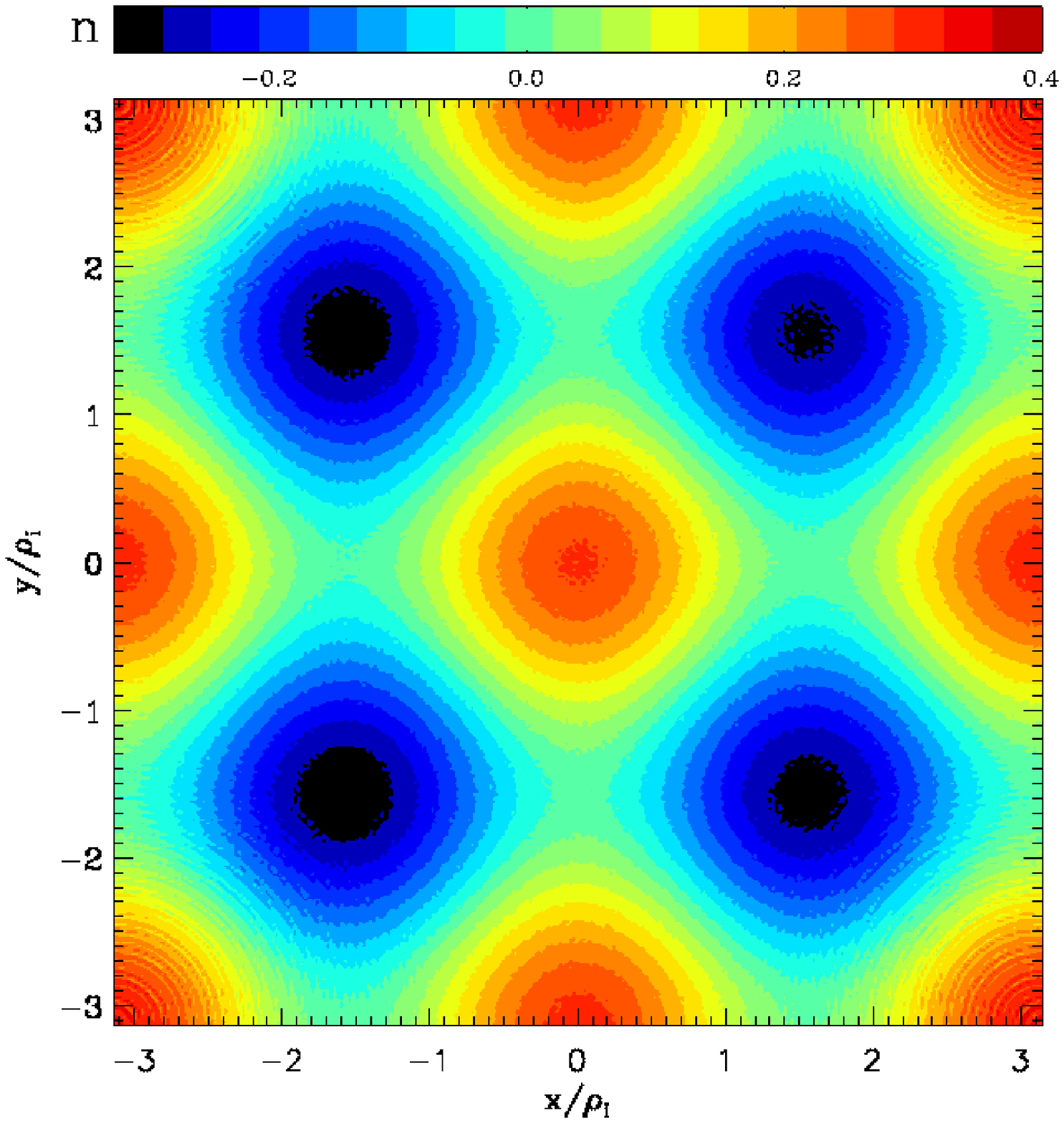}
  \end{minipage}
  \begin{minipage}[!h]{0.2385\textwidth}
  \flushleft\includegraphics[width=1.0\textwidth]{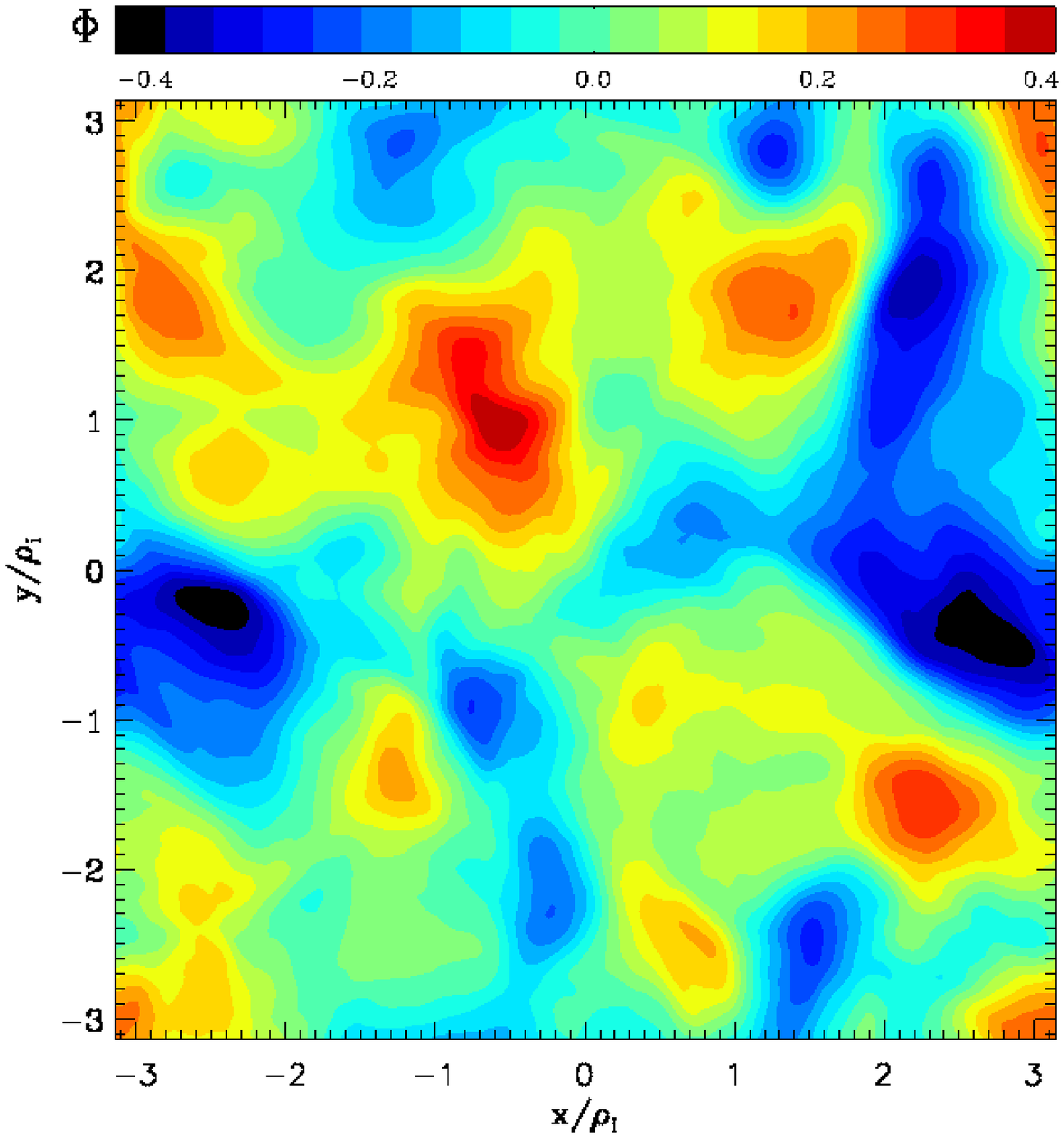}
  \end{minipage}
\caption{Contour plots of the initial condition $n_1$ (left) and of a later turbulent state for $\phi_1$ (right).}
\label{fig:fig0}
\end{figure}

\begin{figure}[!h]
  \begin{minipage}[!h]{0.49\textwidth}
  \flushleft\includegraphics[width=1.0\textwidth]{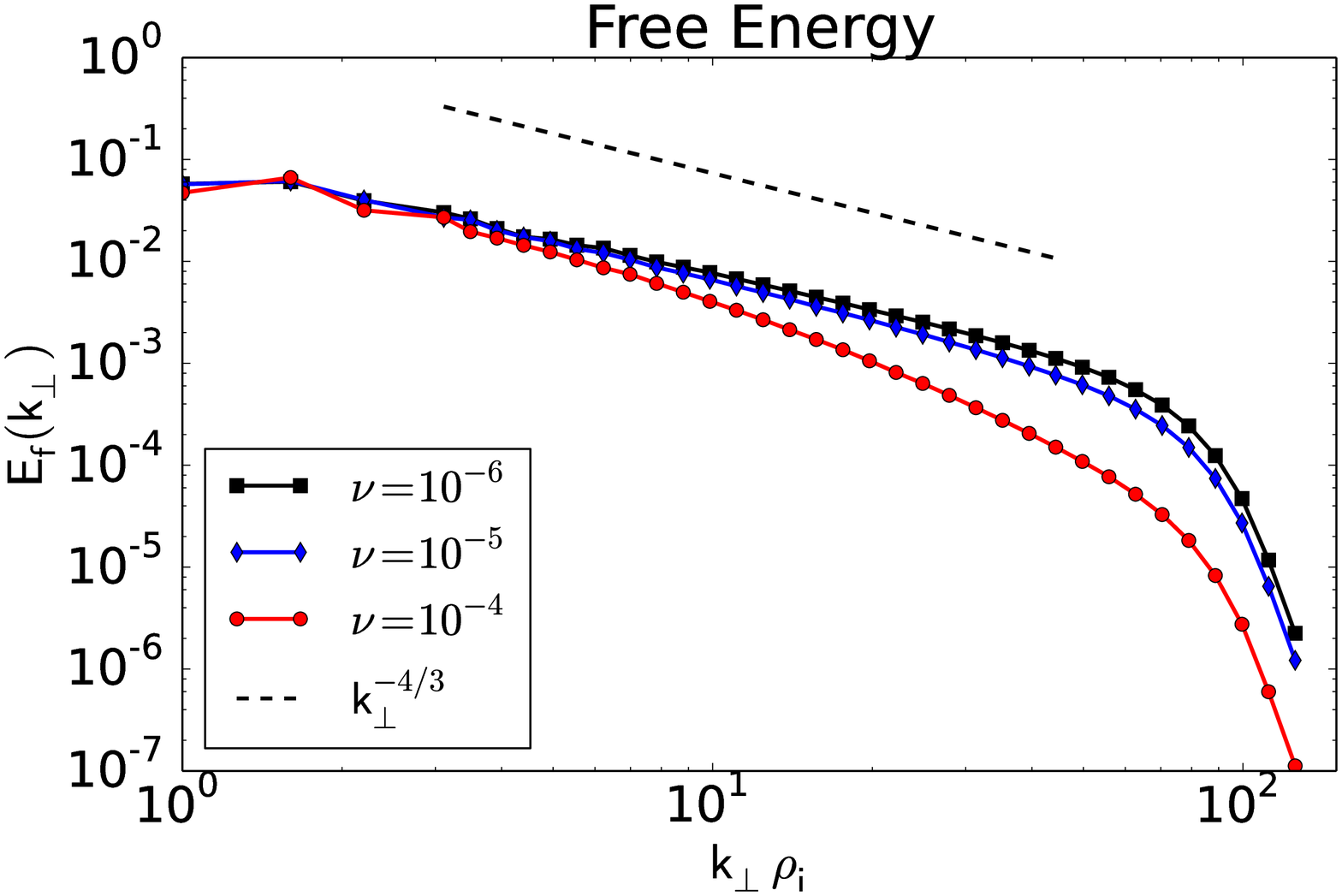}
  \end{minipage}\\
  \begin{minipage}[!h]{0.49\textwidth}
  \flushleft\includegraphics[width=1.0\textwidth]{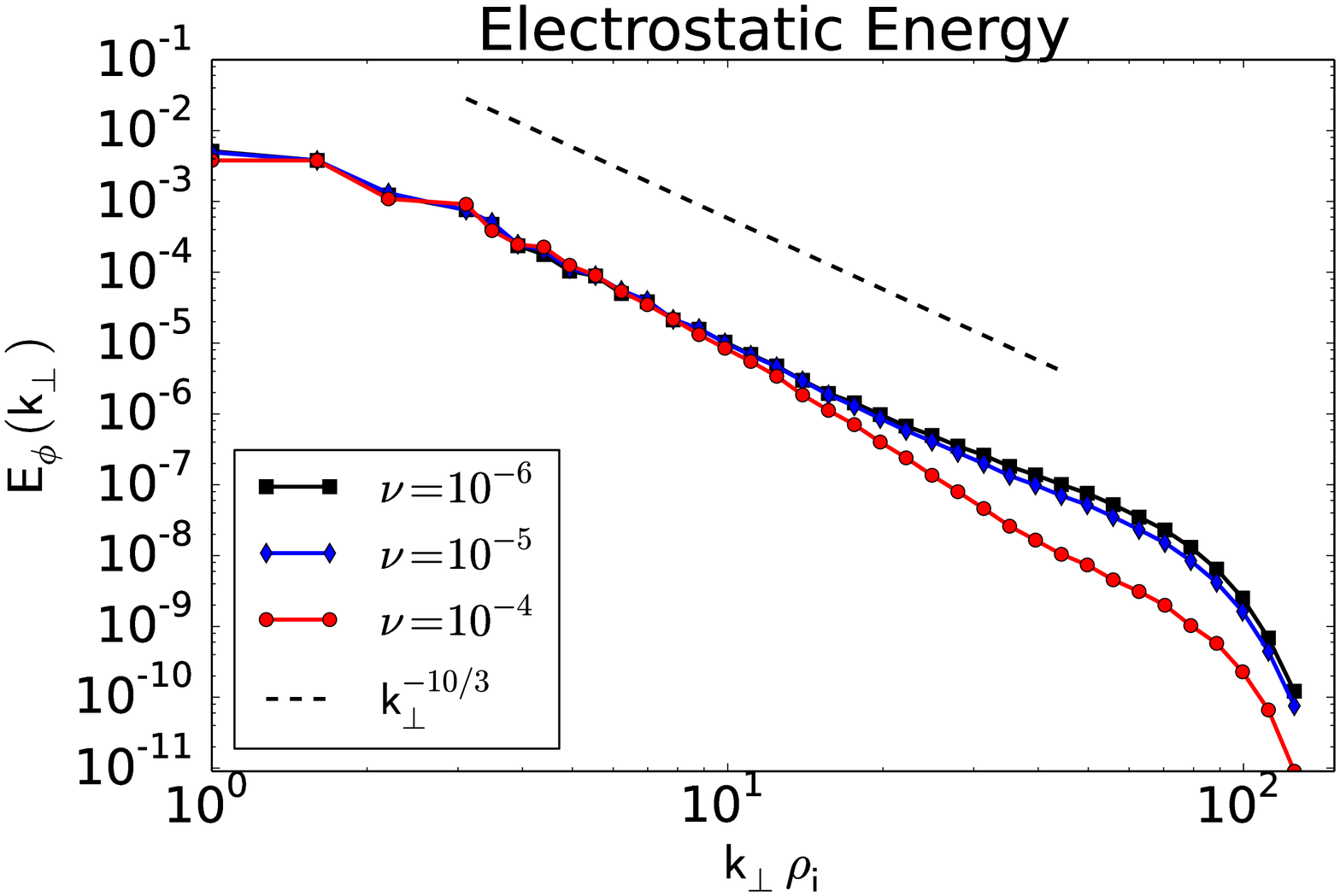}
  \end{minipage}
\caption{$E_f(k_\perp)$ (top) and $E_\phi(k_\perp)$ (bottom) for three different collision frequencies: $\nu=10^{-6}$, $10^{-5}$, and $10^{-4}$ (black squares, blue diamonds, and red circles, respectively). The dashed lines represent the weakly collisional theory.}
\label{fig:fig1}
\end{figure}

\begin{figure}[!h]
  \begin{minipage}[!h]{0.5\textwidth}
  \flushleft\includegraphics[width=1.0\textwidth]{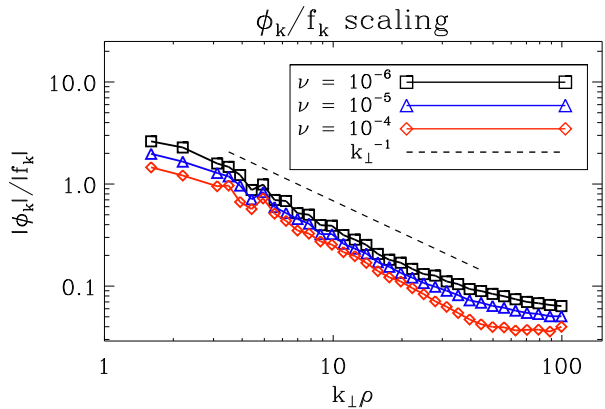}
  \end{minipage}
\caption{Scaling relation $\phi_k/f_k$ for three different collision frequencies: $\nu=10^{-6}$, $10^{-5}$, and $10^{-4}$ (black squares, blue triangles, and red diamonds, respectively). The dashed line corresponds to the scaling $k_\perp^{-1}$ predicted by Ref.~\citep{PlunkJFM2010}.}
\label{fig:fig2}
\end{figure}

\begin{figure}[!h]
  \begin{minipage}[!h]{0.5\textwidth}
  \center\includegraphics[width=1.0\textwidth]{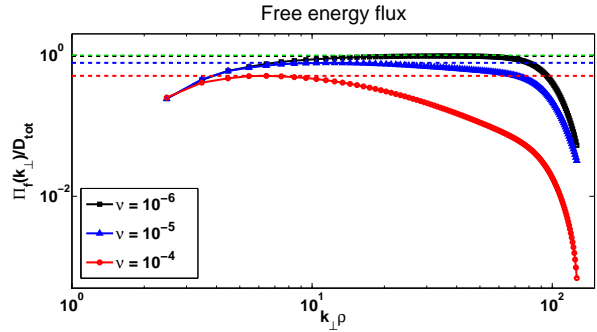}
  \end{minipage}
\caption{Normalized free energy flux $\Pi_f(k_\perp)/D_{\rm tot}$ for $\nu=10^{-6}$, $10^{-5}$, and $10^{-4}$ (black squares, blue triangles, and red circles, respectively). The green dashed line corresponds to unity, while the other colored dashed lines corresponds to the maximum of the respective flux.}
\label{fig:fig3}
\end{figure}

In order to test these ideas, the nonlinear GK equations, Eqs.~(\ref{eq:GK-Vlasov}) and (\ref{eq:GK-Poisson}), are solved by means of direct numerical simulations with the GENE code in a 4D phase space ($x$, $y$, $v_\|$, $\mu$). The size of the domain in real space is $L_x=L_y=2\pi\rho$, while the velocity-space domain is bounded by $-3v_T\leq v_\|\leq +3v_T$ and $0\leq\mu\leq9T_0/B_0$, where $v_T=\sqrt{2T_0/m}$ is the thermal velocity. We use ($256$, $128$, $32$, $96$) points in our 4D phase-space and a linearized Landau collision operator acting on $F_1/F_0$ in gyrocenter coordinates~\citep{Merz_PhD2009, AbelPOP2008, BarnesPOP2009, Doerk_PhD2012} is adopted for $\langle C^L[F_1]\rangle$. Note that, since we are using a Fourier representation, the resolution we have indicated as ($256$, $128$) in the Fourier modes actually corresponds to ($256$, $256$) (fully dealiased) grid points in real-space coordinates.
We remind the reader that there are no gradients in the background quantities, $n_0$, $T_0$, and $B_0$. Thus the system is initialized with an appropriate perturbed distribution function $F_1$, and then it may freely evolve. In the simulations $B_0=1$ and $T_0=1$, so the $\mu$-grid resolution is high enough to account for the largest $k_\perp$ modes and thus for the expected nonlinear phase mixing argument $l\sim l_v$ to hold also in the numerical framework.
We have chosen the same initial condition as given in Ref.~\citep{TatsunoPRL2009}, i.e., a Maxwellian in velocity space and a sum of $k_x=2$ and $k_y=2$ cosines with a small-amplitude white noise $\chi$ on all Fourier modes:
%%%%%%%%%%%%%%%%%%%%%%%%%%%%%%%%%%%%%%%%%%%%%%%%%%%%%
\[ F_1(x,y,v_\|,\mu;t=0) = \]
\[ \digamma_0[\cos(2x/\rho)+\cos(2y/\rho)+\epsilon\chi(x,y)]F_0(v_\|,\mu)\,, \]
%%%%%%%%%%%%%%%%%%%%%%%%%%%%%%%%%%%%%%%%%%%%%%%%%%%%%
where $\digamma_0$ is a constant, $\epsilon$ is the (small) amplitude of the white noise with respect to the cosine functions and $F_0$ is the background Maxwellian. With time, the system evolves into a turbulent state (see Fig.~\ref{fig:fig0}).
%================================
%================================
Moreover, in order to avoid energy pile-up at the very end of the spectrum (high-$k_\perp$), an 8th-order $k_\perp$-hyperdiffusion operator $H_\perp[F_1]=-a_\perp(k_\perp/k_0)^8F_1$ is added on the RHS of Eq.~(\ref{eq:GK-Vlasov}). Note that, due to the very high order of the hyperdiffusion and to an appropriate choice of $k_0$, $H_\perp[F_1]$ will be relevant only for very high $k_\perp$ and thus negligible in the $k_\perp$-range of interest for the spectra (e.g., $a_\perp=0.5$ and $k_0=96$ in our runs). We remark that, since we are dealing with decaying turbulence, the spectra must be normalized appropriately at each time step (e.g., w.r.t.~the amplitude of the decaying quantity) and then time-averaged over a (collisionality dependent) interval [$t_a$, $t_b$]. In particular, the parameter $t_b$ can be chosen as the maximum simulation time, $t_{\rm max}=120$, for every collision frequency $\nu$, provided that we are in a state of almost completely decayed turbulence (so the spectra do not change if we pass from $t_b=t_{\rm max}$ to $t_b=t_{\rm max}-\Delta t$, with $\Delta t$ sufficiently small). The parameter $t_a$ is collisionality dependent, since it must be choosen in a way such that the turbulence is fully developed, which actually depends on $\nu$. For a quantity $A_{ijs}\equiv A(k_{x,i},k_{y,j},t_s)$, the spectrum $E_A(n)=E_A(k_{\perp,n})$ is defined by
%%%%%%%%%%%%%%%%%%%%%%%%%%%%%%%%%%%%%%%%%%%%%%%%%%%%% 
\[
 E_A(n) \equiv \widetilde{\sum}_{(i,j)\in n}E_A(i,j) = \widetilde{\sum_{i,j}}\sum_s\frac{w_s|A_{ijs}|^2}{\sum_{i,j}|A_{ijs}|^2}\,,
\]
%%%%%%%%%%%%%%%%%%%%%%%%%%%%%%%%%%%%%%%%%%%%%%%%%%%%% 
where $\widetilde{\sum}_{(i,j)\in n}$ is the ``ring average'' over the $n$-th shell, and, since GENE uses adaptive time steps, $w_s$ are the corresponding ``time weights''. 
In Fig.~\ref{fig:fig1} we report the numerical results for the free energy spectrum $E_f(k_\perp)$ and the electrostatic energy spectrum $E_\phi(k_\perp)$ from simulations with collision frequencies of $\nu=10^{-6}$, $10^{-5}$, and $10^{-4}$ (black, blue, and red curves, respectively). In order to make more clear how collisional the turbulence is, according to Refs.~\citep{TatsunoPRL2009,TatsunoJPFR2010}, these three cases correspond to a Dorland number of $D=40$, $527$ and $5438$, respectively for $\nu=10^{-4}$, $10^{-5}$ and $10^{-6}$. In particular, this justifies the use of hyperdiffusion, since in the very high Dorland number regime the spectral cutoff $k_{\perp,{\rm c}}\rho\propto D^{3/5}$ would exceed the resolution, thus we need to impose an artificial cutoff. Moreover, we used the same hyperdiffusion parameters for all the cases in order to make them comparable.
For these cases, the initial time for the average was chosen to be $t_a=40$, $38$, and $35$ for $\nu=10^{-6}$, $10^{-5}$, and $10^{-4}$, respectively. As we can see, the spectra become steeper with increasing collision frequency $\nu$. This is indeed what can be expected qualitatively from Eq.~(\ref{eq:spectra}): good agreement with the standard theory, Eq.~(\ref{eq:spectra_0}), for low collisionality, and a steepening of the slopes with increasing $\nu$.
%================================
%================================
\\
We would like to stress that, even though the exponential part of the spectrum becomes more important with increasing collisionality, up to $\nu\sim10^{-4}$ the power law part is still dominant and the spectrum can be described by a pure power law for an order of magnitude to a very good approximation (see Fig.~\ref{fig:fig1}).
An overview of the exponents, $\alpha_f$ and $\alpha_\phi$, as inferred from the simulation results shown in Fig.~\ref{fig:fig1} via a linear fit is presented in Table~\ref{tab:tab2}. The uncertainties are estimated from the variation of the exponents due to the choice of the $k_\perp$-range of fitting, the time window of average and the number of bins for the ring average.
%................................
\begin{table}[!h]
 \center
 \begin{tabular}{|c|cc|}
  \toprule
   %\hline
   $\boldsymbol{\nu}$ & $\alpha_f$ & $\alpha_\phi$\\
   \hline 
    10$^{-6}$ & 1.35$\pm$0.05 & 3.45$\pm$0.10\\
    10$^{-5}$ & 1.50$\pm$0.05 & 3.75$\pm$0.10\\
    10$^{-4}$ & 2.05$\pm$0.15 & 4.45$\pm$0.25\\
   %\hline
   \toprule
 \end{tabular}
   \caption{Spectral exponents extracted from the simulations shown in Fig.\ref{fig:fig1}.}
   \label{tab:tab2}
\end{table}
%................................
\\
Numerical simulations can also be used to check the hypotheses used in the theory (Sec.~\ref{sec:scalings}). In particular, we are going to test three fundamental features of the theory: (i) the scaling relation between $\phi_k$ and $f_k$, (ii) the free energy flux $\Pi_f$, and (iii) the locality of the energy cascade. 
The first assumption, i.e., the scaling relation in Eq.~(\ref{eq:phi-f_scal}), which reduces to the standard scaling (\ref{eq:phi-f_scal_0}) in the low collisionality limit ($\theta_\nu\to0$), is displayed in Fig.~\ref{fig:fig2} for the three cases $\nu=10^{-6}$, $10^{-5}$, and $10^{-4}$ of Fig.~\ref{fig:fig1}. Note that in Fig.\ref{fig:fig2} we have introduced a shift in magnitude of the relation for the three cases for the sake of clarity. The simulations thus demonstrate that the scaling relation between $\phi_k$ and $f_k$ becomes steeper than $k_\perp^{-1}$ with increasing collisionality, although $\theta_\nu\ll1$ is a only a relatively small correction (e.g., $\theta_\nu\simeq0.15\pm0.05$ for $\nu=10^{-4}$). This was expected, since we are still in a regime in which the collisions do not affect the nonlinear phase mixing argument, $l_v\sim l$, but they just avoid to form a lot of small scale structures in $v_\perp$-space.
%================================
%================================
\\
In Fig.~\ref{fig:fig3}, the second assumption, regarding the flux of free energy, Eq.~(\ref{eq:PIf_hp}), is investigated. Here, the flux is normalized to the total dissipation. The non-constancy of the flux reflects the fact that dissipation is actually effective at all scales, since for purely conservative spectral energy transfer (i.e., the standard picture of the inertial range), one expects $\Pi_f/D_{\rm tot}=1$ for an extended region in $k$-space, until the dissipation range begins and an exponential fall-off appears. However, this is only approximately the case even at the lowest collisionality, $\nu=10^{-6}$, for which $(\Pi_f/D_{\rm tot})_{\rm max}\sim0.95$. Meanwhile, we obtain $(\Pi_f/D_{\rm tot})_{\rm max}\sim0.75$ for $\nu=10^{-5}$ and $(\Pi_f/D_{\rm tot})_{\rm max}\sim0.5$ for $\nu=10^{-4}$, after which the flux is not even constant (it can be considered a power law, as a first approximation). This behavior clearly deviates from the standard picture of an inertial range.
%================================
%================================
\\
Moreover, the differences between the assumptions made in the standard theory and the fluxes in Fig.~\ref{fig:fig3} is way more evident than the deviations found in the scalings (Fig.~\ref{fig:fig2}). In other words, $\delta_\nu$ is deviating from zero more rapidly due to the multiscale dissipation: e.g., we estimate $\delta_\nu\simeq1.0\pm0.2$ for $\nu=10^{-4}$. Again, this means that the weaker point is to assume the standard picture of an inertial range bridging the injection and the disspation scales: this is something that was already observed in gyrokinetics for other type of systems (see Refs.~\citep{TeacaPRL2012,HatchPRL2013,TeacaPOP2014}).
As a final remark on these two assumptions, we note that not only qualitative agreement is found: the values of $\theta_\nu$ and $\delta_\nu$ estimated from simulations are also in quantitative agreement with the fitted exponents of the spectra. In fact, for instance, we found $\alpha_f\simeq2.05\pm0.15$ and $\alpha_\phi\simeq4.45\pm0.25$ for the $\nu=10^{-4}$ case (Table~\ref{tab:tab2}), while the values of $\theta_\nu$ and $\delta_\nu$ extracted from the simulations (Figs.~\ref{fig:fig2} and \ref{fig:fig3}) predict $\alpha_f\approx1.90\pm0.15$ and $\alpha_\phi\approx4.20\pm0.25$. Considering the approximations made and the uncertainties, this is a good agreement.\\
We finally test the third assumption, i.e., the locality of the energy cascade. This feature can be checked by looking at the shell-to-shell transfer, $T_n^m$, i.e., the energy exchange between the $k_\perp$ shells. Defining the shells as $\{k_n\}_{n=1,2,\dots,N}$, where we fix three parameters, $k_a$ and $k_b$ and $A$, such that $k_1=k_a$, $k_2=k_a+k_b$, and $k_n=(k_a+k_b)2^{(n-2)/A}=k_22^{(n-2)/A}$ for $n=2,\dots,N$. In the following, $k_a=k_b=4$ and $A=5$ will be adopted. Note that $T_n^m$ is the discrete version of ${\cal T}_f(k',k)$. 
%================================
\begin{figure}[!h]
  \begin{minipage}[!h]{0.2385\textwidth}
  \flushleft\includegraphics[width=1.05\textwidth]{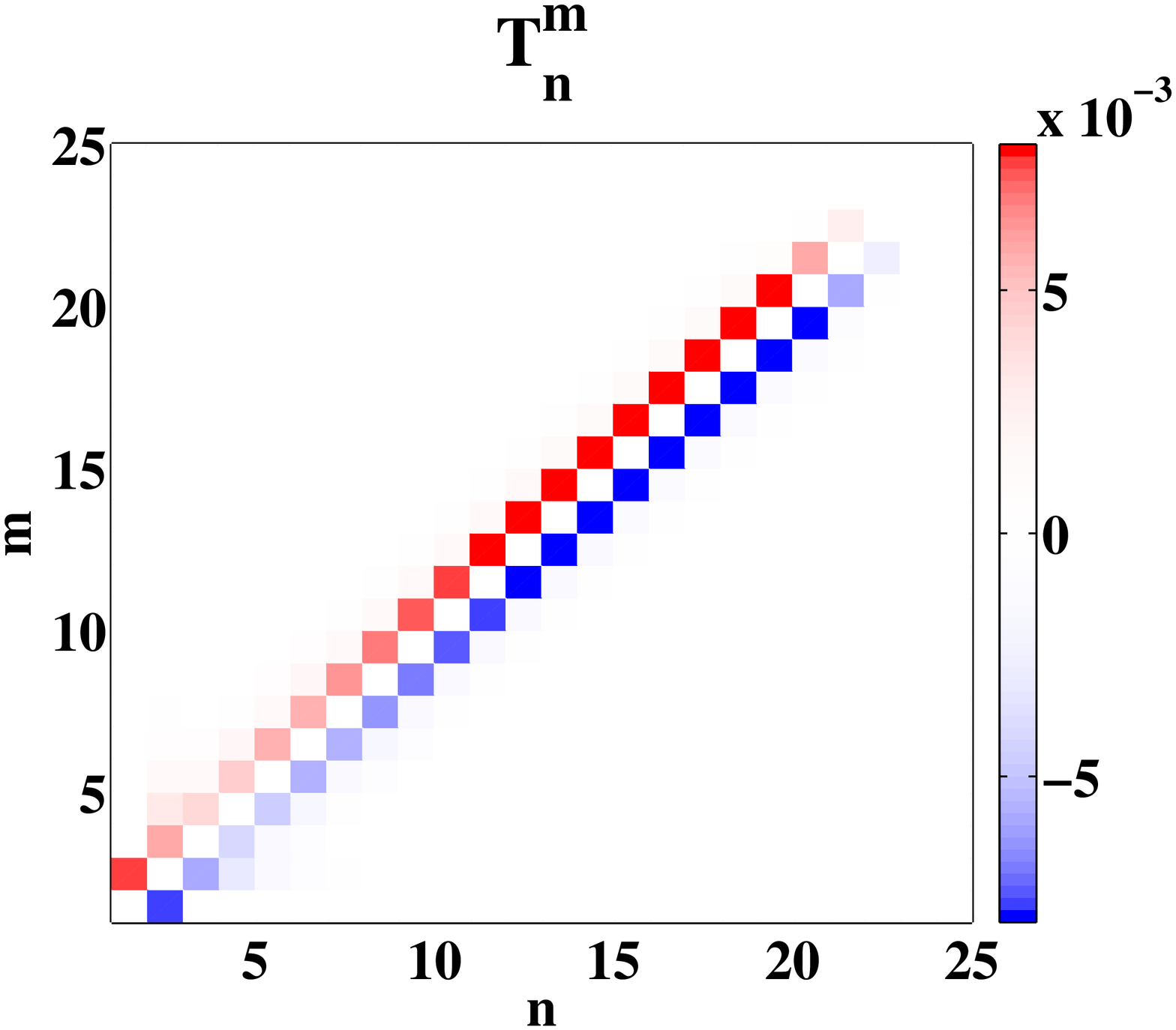}
  \end{minipage}
  \begin{minipage}[!h]{0.2385\textwidth}
  \flushleft\includegraphics[width=1.05\textwidth]{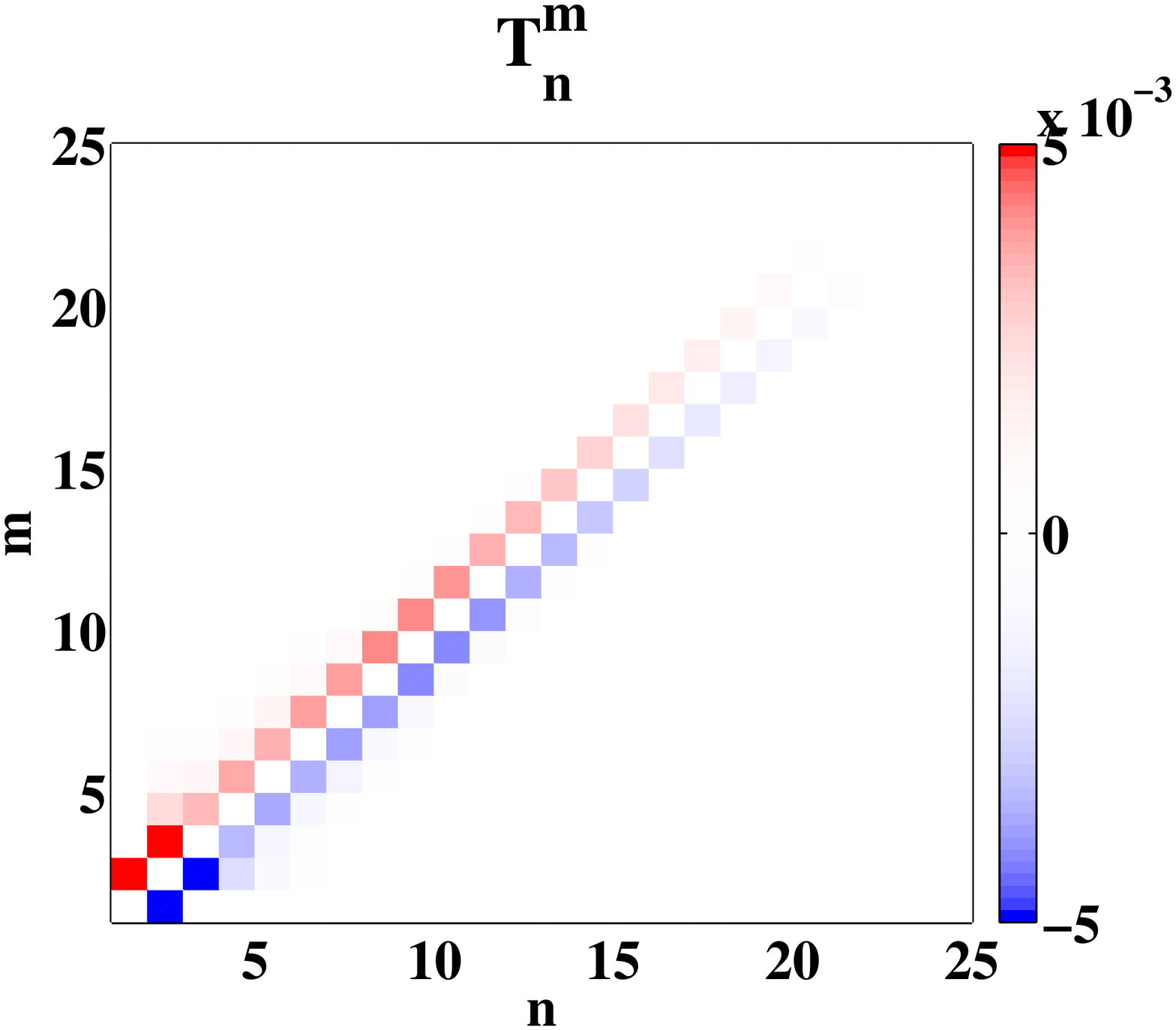}
  \end{minipage}
\caption{Shell-to-shell transfer $T_n^m$ for $\nu=10^{-6}$ (right) and for $\nu=10^{-4}$ (left).}
\label{fig:fig4}
\end{figure} 
%================================
The shell-to-shell transfer for $\nu=10^{-6}$ and $\nu=10^{-4}$ is shown in Fig.~\ref{fig:fig4} (left and right panel, respectively): we can see that the energy transfer is very local. I.e., the free energy exchange is relevant only between neighboring shells, thus verifying the locality assumption. Moreover, the antisymmetry of $T_n^m$ is immediately evident from the plot, which means it is a non-dissipative term that only transfers energy between the modes, as to be expected. From Fig.~\ref{fig:fig4}, we also recognize the features already seen in Fig.~\ref{fig:fig3} about the free energy flux $\Pi_f(k_\perp)$, i.e., the non-constancy of the flux for $\nu=10^{-4}$, denoted by the softening of the colors, in contrast to its nearly-constant behavior for the $\nu=10^{-6}$ case.

%*****************%
%   CONCLUSIONS   %
%*****************%
\section{Conclusions and discussion}\label{sec:end}

We have presented a theoretical and numerical study of freely decaying electrostatic turbulence in the framework of collisional gyrokinetic theory of magnetized plasmas. A reduced 4D (2D2V) phase space ($x$, $y$, $v_\|$, $\mu$) is considered: the 2D real space is perpendicular to the background magnetic field, thus avoiding parallel effects such as the Landau damping to be effective. No background gradients were considered, making the system as simple as possible, and thus focusing on the two effects of interest: the E$\times$B nonlinearity and collisions. The nonlinear term introduces a perpendicular (nonlinear) phase mixing~\citep{SchekochihinPPCF2008,PlunkJFM2010}, causing the perturbed part of the distribution function to develop structures in $v_\perp$-space which are related to those in real space (or, equivalently, to $k_\perp$). In particular, the higher the $k_\perp$ is, the finer those $v_\perp$-structures are. 
However, the relation between $v_\perp$-scales and $k_\perp$ is affected by the collisionality of the system, which in practice has the function of limiting the finest $v_\perp$-scales. Then, due to the relation between $v_\perp$ scales and real-space scales, this has a direct influence on the $k_\perp$ scalings between $\phi_k$ and $f_k$. This was indeed shown by means of direct numerical simulations, even if it remains a small effect, provided that the collisionality is not too high. %to invalidate the nonlinear phase mixing argument. 
In addition, allowing for the presence of multiscale dissipation in sub-Larmor scale fluctuations has immediate consequences on one of the very fundamental assumptions made by standard Kolmogorov-like theories, i.e., on the existence of an inertial range. In fact, regarding this point, we have shown that dissipation occurs at all scales, regardless of the collisionality regime and, moreover, that for intermediate collisionality the free energy flux is not even nearly constant anymore. This leaves a very important fingerprint on the spectra, making in fact the power law collisionality dependent and thus allowing for non-universal power laws.

Despite the relative simplicity of the system under study here, it seems plausible that these results can be generalized and applied to more complicated systems. For instance, it is reasonable to expect that also 3D GK turbulence in toroidal fusion devices can exhibit variable power law scalings depending on the parameter settings. Such behaviour has indeed been observed before, and the present work offers a possible explanation. Similarly, 3D GK turbulence simulations applied to the solar wind dissipation range display non-universal power law scalings as well as exponential corrections. Again, the present work may provide a key to the understanding of this effect. Follow-up studies will have to clarify if these conjectures are correct.

%*******************%
%   Aknowledments   %
%*******************%
\subsection*{Acknowledgments}

We would like to thank H. Doerk, V. Bratanov, H. Sceats, G. G. Plunk and A. A. Schekochihin for very useful discussions.
The research leading to these results has received funding from the European Research Council under the European Union’s Seventh Framework Programme (FP7/2007-2013)/ERC Grant Agreement No. 277870. This project has received funding from the Euratom research and training programme 2014-2018.

%******************%
%   Bibliography   %
%******************%


\begin{thebibliography}{99.}

\bibitem{Frisch1995} 
U. Frisch, {\em Turbulence}, Cambridge University Press (1995).

\bibitem{FalkovichPT2006} 
G. Falkovich and K. R. Sreenivasan, Phys. Today {\bf59}(4), 43 (2006).

\bibitem{FriemannPOF1982} 
E. A. Friemann and L. Chen, Phys. Fluids {\bf25}, 502 (1982)

\bibitem{BrizardRMP2007} 
A. J. Brizard and T. S. Hahm, Rev. Mod. Phys. {\bf79}, 421 (2007).

\bibitem{KrommesARFM2012} 
J. A. Krommes, Ann. Rev. Fluid Mech. {\bf44}, 175 (2012).

\bibitem{SchekochihinAPJS2009} 
A. A. Schekochihin, S. C. Cowley, W. Dorland, G. W. Hammett, G. G. Howes, E. Quataert, T. Tatsuno, ApJS {\bf 182}, 310 (2009).

\bibitem{BanonNavarroPRL2011} 
A. Ba\~n\'on Navarro, P. Morel, M. Albrecht-Marc, D. Carati, F. Merz, T. G\"orler, F. Jenko, Phys. Rev. Lett. {\bf106}, 055001 (2011).

\bibitem{SchekochihinPPCF2008} 
A. A. Schekochihin, S. C. Cowley, W. Dorland, G. W. Hammett, G. G. Howes, G. G. Plunk, E. Quataert, T. Tatsuno, Plasma Phys. Control. Fusion {\bf 50}, 124024 (2008).

\bibitem{PlunkJFM2010} 
G. G. Plunk, S. C. Cowley, A. A. Schekochihin, T. Tatsuno, J. Fluid Mech. {\bf 664}, 407 (2010).

\bibitem{TatsunoPRL2009} 
T. Tatsuno, W. Dorland, A. A. Schekochihin, G. G. Plunk, M. Barnes, S. C. Cowley, G. G. Howes, Phys. Rev. Lett. {\bf 103}, 015003 (2009).

\bibitem{TatsunoJPFR2010} 
T. Tatsuno, M. Barnes, S. C. Cowley, G. G. Howes, R. Numata, G. G. Plunk, A. A. Schekochihin, J. Plasma Fusion Res. SERIES {\bf9}, 509 (2010).

\bibitem{JenkoPOP2000} 
F. Jenko, W. Dorland, M. Kotschenreuther, B. N. Rogers, Phys. Plasmas {\bf7}, 1904 (2000).

\bibitem{DannertPOP2005} 
T. Dannert and F. Jenko, Phys. Plasmas {\bf12}, 072309 (2005).

\bibitem{GoerlerJCP2011} 
T. G\"orler, X. Lapillonne, S. Brunner, T. Dannert, F. Jenko, F. Merz, D. Told, J. Comput. Phys. {\bf230}, 7053 (2011).

\bibitem{LandauJP_1946} 
L. D. Landau, J. Phys. (USSR) {\bf10}, 25 (1946).

\bibitem{HammettPFB-4_1992} 
G. W. Hammett, W. Dorland, F. W. Perkins, Phys. Fluids B {\bf4}, 2052 (1992).

\bibitem{BanonNavarroPOP2011} 
A. Ba\~n\'on Navarro, P. Morel, M. Albrecht-Marc, D. Carati, F. Merz, T. G\"orler, F. Jenko, Phys. Plasmas {\bf18}, 092303 (2011).

\bibitem{TeacaPRL2012} 
B. Teaca, A. Ba\~n\'on Navarro, F. Jenko, S. Brunner, L. Villard, Phys. Rev. Lett. {\bf109}, 235003 (2012).

\bibitem{HatchPRL2013} 
D. R. Hatch, F. Jenko, A. Ba\~n\'on Navarro, V. Bratanov, Phys. Rev. Lett. {\bf111}, 175001 (2013).

\bibitem{TeacaPOP2014} 
B. Teaca, A. Ba\~n\'on Navarro, F. Jenko, ``The energetic coupling of scales in gyrokinetic plasma turbulence'', Phys. Plasmas (submitted).

\bibitem{lv_note} 
If $\delta v_\perp$ is the typical velocity scale length between the fluctuations developed by the distribution function $F_1$ in $v_\perp$-space (see later in the text), then we can define $l_v\sim (\delta v_\perp/v_T)\rho$ (see, e.g., Ref.~\citep{SchekochihinPPCF2008}).

\bibitem{Merz_PhD2009} F. Merz, Ph.D. thesis, Universit\"at M\"unster, 2008.

\bibitem{AbelPOP2008} I. G. Abel, M. Barnes, S. C. Cowley, W. Dorland, A. A. Schekochihin, Phys. Plasmas {\bf15}, 122509 (2008).

\bibitem{BarnesPOP2009} M. Barnes, I. G. Abel, T. Tatsuno, A. A. Schekochihin, S. C. Cowley, W. Dorland, Phys. Plasmas {\bf16}, 072107 (2009).

\bibitem{Doerk_PhD2012} 
H. Doerk, Ph.D. thesis, Universit\"at Ulm, 2012.


\end{thebibliography}
\end{document}